\input{aipcheck}

\documentclass[
    ,final            % use final for the camera ready runs
  ]
  {aipproc}

\layoutstyle{6x9}

\begin{document}

\title{Redshift and spatial distribution of the intermediate gamma-ray bursts}


\author{I.~Horv\'ath}{
  address={Dept. of Physics, Bolyai Military University, POB 15, 1581 Budapest, Hungary.} 
}

\author{Z.~Bagoly}{
  address={Dept. of Physics of Complex Systems, E\"otv\"os Univ., P\'azm\'any P. s. 1/A, 1117 Budapest, Hungary.}
  ,altaddress={Dept. of Physics, Bolyai Military University, POB 15, 1581 Budapest, Hungary.} 
}
\author{A.~de~Ugarte~Postigo}{
  address={Dark Cosmology Centre, Niels Bohr Inst., Juliane Maries Vej 30, Copenhage \O, 2100, Denmark}
  }

\author{L.~G.~Balazs}{
  address={Konkoly Observatory, 1525 Budapest, POB 67, Hungary.}
}

\author{P.~Veres}{
  address={Dept. of Physics, Bolyai Military University, POB 15, 1581 Budapest, Hungary.}
  ,altaddress={Dept. of Physics of Complex Systems, E\"otv\"os Univ., P\'azm\'any P. s. 1/A, 1117 Budapest, Hungary.} 
}

\begin{abstract}

One of the most important task of the Gamma-Ray Burst field is the 
classification of the bursts. Many researches have proven the 
existence of the third kind (intermediate duration) of GRBs in 
the BATSE data. Recent works have analyzed BeppoSax and 
Swift observations and can also identify three types of GRBs 
in the data sets. However, the class memberships are 
probabilistic we have enough observed redshifts to calculate 
the redshift and spatial distribution of the intermediate GRBs.
They are significantly farther than the short bursts and seems to
be closer than the long ones.
  \end{abstract}

\maketitle

%%%%%%%%%%%%%%%%%%%%%%%%%%%%%%%%%%%%%%%%%%%%
%% MAINMATTER
%%%%%%%%%%%%%%%%%%%%%%%%%%%%%%%%%%%%%%%%%%%%

\section{Introduction}
In the Third BATSE Catalog  --- using uni- and
multi-variate analyses --- \citet{hor98} and \citet{muk98} found a third
type of GRBs. Later several papers
\citep{hak00,bala01,rm02,hor02,hak03,bor04,hor06,chat07} confirmed the
existence of this third ("intermediate" in duration) group in
the same database.
Here we classify the 237 GRBs from the
Swift first BAT catalog \citep{sak08}.
Using this result we calculated the redshift distributions
for the classes.

\section{Classification of \textit{Swift} GRBs}

The probability distribution of
the logarithm of durations ($x$) can be well fitted by Gaussian
distributions, if we restrict ourselves to the short and long
GRBs \citep{hor98}. We assume the same also for the $y$
coordinate. With this assumption we obtain, for a certain $l$-th
class of GRBs,
$$
p(x,y|l)  =
 \frac{1}{2 \pi \sigma_x \sigma_y
\sqrt{1-r^2}} \times \;\;\;\;\;\;\;\;\;\;\;\;\;\;\;\;\;\;\;\;
\;\;\;\;\;\;\;\;\;\;\;\;$$
\begin{equation} \label{gauss}
\exp\left[-\frac{1}{2(1-r^2)}
\left(\frac{(x-a_x)^2}{\sigma_x^2} + \frac{(y-a_y)^2}{\sigma_y^2}
- \frac{2r(x-a_x)(y-a_y)} {\sigma_x \sigma_y}\right)\right] \;
\end{equation}
where  $a_x$, $a_y$ are the means, $\sigma_x$, $\sigma_y$ are
the dispersions, and $r$ is the correlation coefficient. 
Hence, a certain class is defined by five independent
parameters, $a_x$, $a_y$, $\sigma_x$, $\sigma_y$, and $r$, which are
different for different $l$. If we have $k$ classes, then we
have $(6k - 1)$ independent parameters (constants), because any
class is given by the five parameters of Eq.(\ref{gauss}) and
the weight $p_l$ of the class. One weight is not independent,
because $\sum \limits_{l=1}^{k} p_l = 1$. The sum of $k$
functions defined by Eq.(\ref{gauss}) gives the theoretical
function of the fit.

\begin{figure}[h!]
  \includegraphics[height=.4\textheight,width=0.9\textwidth]{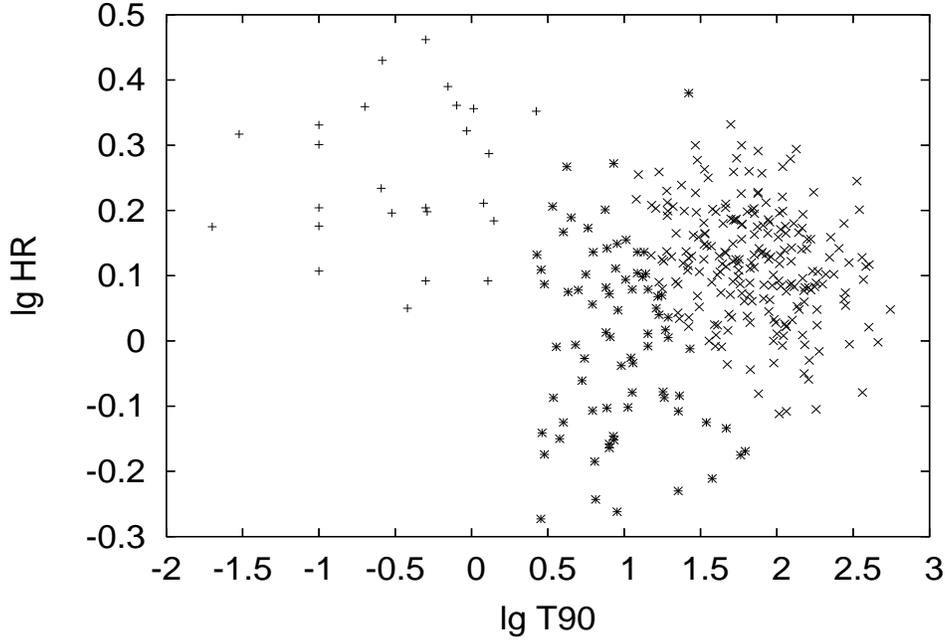}
  \caption{The classification result in the duration ($x$) - hardness ($y$) plane.
Different symbols mark different classes. 
%The green circles are the short bursts the red circles are the long ones the
%blue circles are the intermediate GRBs. The size of the cirle is proportional with the
%probalility of the membership.
}\label{fig:dark}
\end{figure}

The result of the three-Gaussian fit is shown in Figure 1. The best parameters 
were published in Horváth et al. 2010 \citep{hor10}. Moving from k = 2 to k = 3 the 
number of parameters m increases by 6 (from 11 to 17) and 
$L_{max}$ grows from 506.6 to 531.4. The increase in $L_{max}$ by a value of 
25 corresponds to a value of 50 for a $\chi^2_6$   distribution. 
The probability $\chi^2_6$ for 50 is very low ($10^{-8}$), therefore we  conclude that 
the inclusion of a third class  is well justified.
	Moving from k = 3 to k = 4, however, the improvement in 
$L_{max}$ is 3.4 (from 531.4 to 534.8), which can 
happen by chance with a probability of 33.9\%. Hence, the inclusion of 
the fourth class is not justified.

\section{Redshift Distributions}

The cumulative redshift distribution of the three populations is shown in Figure 2.
Redshifts were taken from \cite{ama08}. Only a subset of the classified bursts
had redshift information and we considered bursts where the probability of
belonging to a given population is higher than $97\%$. This means $6$ short,
$9$ intermediate, and $50$ long GRBs. The long and short population
redshift distributions are significantly different ($99.4\%$ significance). 
The intermediate GRBs redshift distribution is clearly between the
short and long redshift distributions, which could mean that they are
further than the short bursts and closer than the long ones.
However, probably owing to the small number of data points the
difference is not significant. We have tried several statistical tests.
None of them showed high significance; the best one
was $92\%$.

\begin{figure}[h!]
  \includegraphics[angle=270,width=0.7\textwidth]{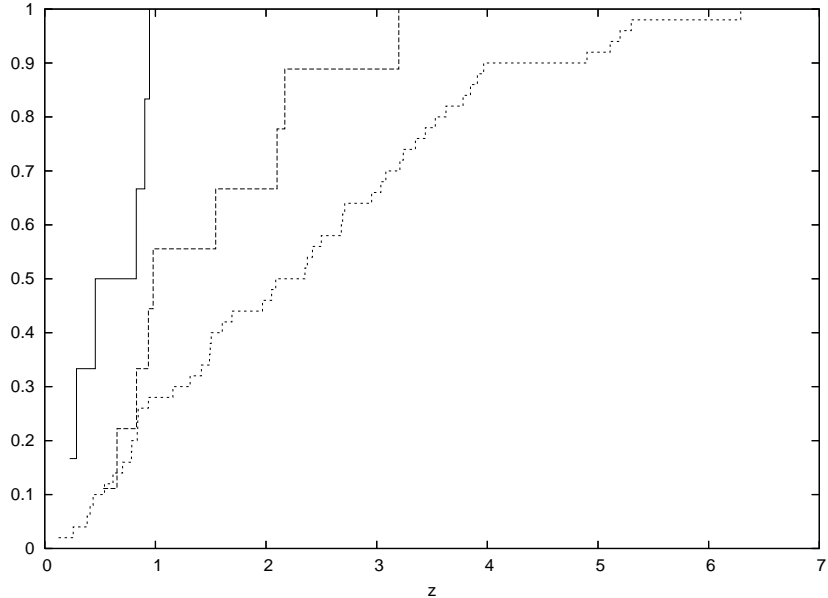}
  \caption{Cumulative redshift distribution of the three classes. 
The continuous line is the short, the dashed line is the intermediate,
and the dotted line is the long population.
   } \label{fig:z}\end{figure}

\section{The 
$E_{\mathrm{peak}}-E_{\mathrm{iso}}$ (Amati) Relation }

Once we classify the bursts, it is also possible to investigate their properties in
the context of the $E_{\mathrm{peak}}-E_{\mathrm{iso}}$ or Amati-relation
\citep{ama08} in the case of bursts with measured redshift. 
%The Amati-relation \citep{ama02} 
%is a correlation between the rest-frame peak energy of the GRB spectrum
%($E_{\mathrm{peak}}$) and the isotropic-equivalent energy release of the
%burst($E_{\mathrm{iso}}$). 
 We have $18$ redshifts from the long
population and $6$ from the intermediate. 
Both groups seem to
follow the same relationship. As the
Amati-relation is not valid for the short population, the intermediate
bursts are more closely related to the long population than to the short class. 
Intermediate bursts do not populate the most energetic regime of the
$E_{\mathrm{peak}}-E_{\mathrm{iso}}$ plane unlike the long bursts (see Figure
3). They tend to have lower isotropic energies compared to the long
population. The small number of data points makes hard to give firm assertions at
this time. Also, there is no significant clustering of intermediate bursts on
this plane.

\begin{figure}[h!]
  \includegraphics[height=.4\textheight]{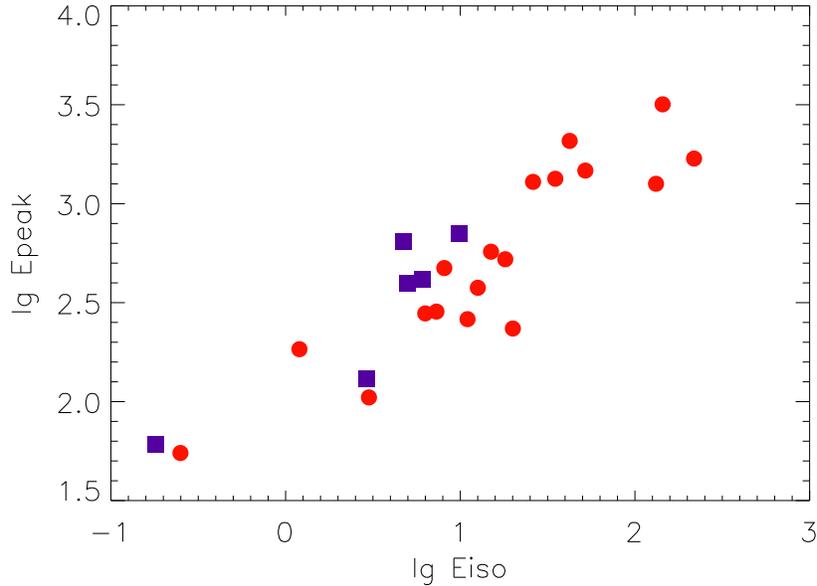}
  \caption{
  $E_{\mathrm{peak}}-E_{\mathrm{iso}}$ relation of the classified
bursts. Circles represent long GRBs, and intermediate bursts are
plotted in square.
The y-axis is the 10-base logarithm of $E_{\mathrm{peak}}$ in keV, and the
x-axis is the 10-base logarithm of $E_{\mathrm{iso}}$ in units of
$10^{52}$ erg/s.  } \label{fig:am}\end{figure}

\begin{theacknowledgments}
This work is partially supported by ASI grant SWIFT I/011/07/0, of the Ministry
of University and Research of Italy (PRIN MIUR 2007TNYZXL), by OTKA K077795, 
by OTKA/NKTH A08-77719 and by A08-77815. The Dark Cosmology Centre is funded by the DNRF.

\end{theacknowledgments}



\bibliographystyle{aipproc}   % if natbib is available

\bibliography{horvath}
\end{document}